\documentclass[twocolumn,twoside,prl,floatfix,letterpaper]{revtex4}

\newif\ifpdf \ifx\pdfoutput\undefined \pdffalse \else \pdftrue \fi
\ifpdf 
  \usepackage[pdftex]{graphicx} 
\else 
  \usepackage{graphicx} 
\fi
\usepackage{amsmath}
\usepackage{amssymb}
\usepackage{fancyhdr}
\usepackage{journals}
\usepackage{color}
\usepackage{rotating}
\usepackage[colorlinks,hyperindex]{hyperref}

\def \Bard {{\sc Bard}}
\def \Vista {{\sc Vista}}
\def \Sleuth {{\sc Sleuth}}
\def \Quaero {{\sc Quaero}}
\def \TurboSim {{\sc TurboSim}}
\def \Pythia {{\sc Pythia}}

\def \MadGraph {{\sc MadGraph}}
\def \MadEvent {{\sc MadEvent}}

\begin{document}

\title{\Bard: Interpreting New Frontier Energy Collider Physics}
\author{Bruce Knuteson}
\homepage{http://mit.fnal.gov/~knuteson/}
\email{knuteson@mit.edu}
\affiliation{MIT}
\author{Stephen Mrenna}
\homepage{http://home.fnal.gov/~mrenna/}
\email{mrenna@fnal.gov}
\affiliation{FNAL}


\begin{abstract}
 No systematic procedure currently exists for inferring the underlying
physics from discrepancies observed in high energy collider data. We
present \Bard, an algorithm designed to facilitate the process of model
construction at the energy frontier. Top-down scans of model parameter
space are discarded in favor of bottom-up diagrammatic explanations of
particular discrepancies, an explanation space that can be exhaustively
searched and conveniently tested with existing analysis tools.
\end{abstract}

\maketitle


\begin{figure}
\includegraphics[width=2.5in,angle=0]{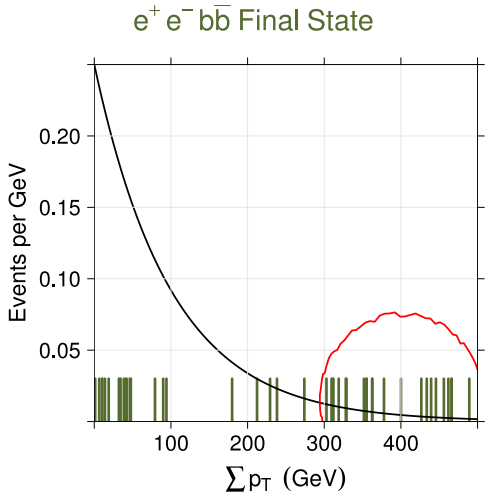}
\caption{A cartoon illustration of \Bard's starting point: an excess (circled in red) in data (individual events shown as tick marks on the horizontal axis) over Standard Model prediction (shown as a continuous distribution) in a particular exclusive final state ($e^+e^-b\bar{b}$) on the tail of the total summed scalar transverse momentum of all objects in the event ($\sum{p_T}$).\label{fig:eebbSleuthCartoon}}
\end{figure}

\begin{figure}
\begin{tabular}{cc}
\includegraphics[width=1.6in]{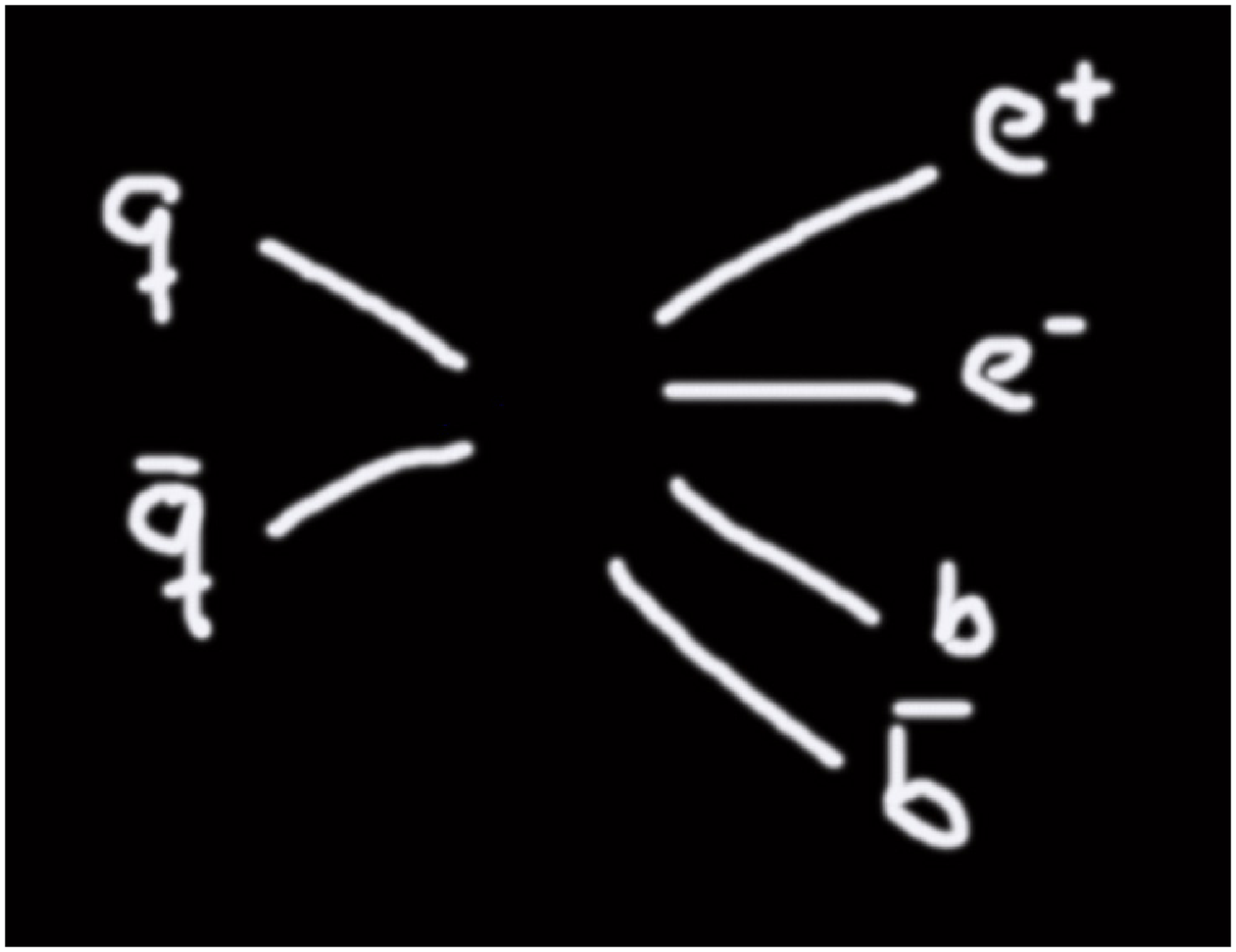} &
\includegraphics[width=1.6in]{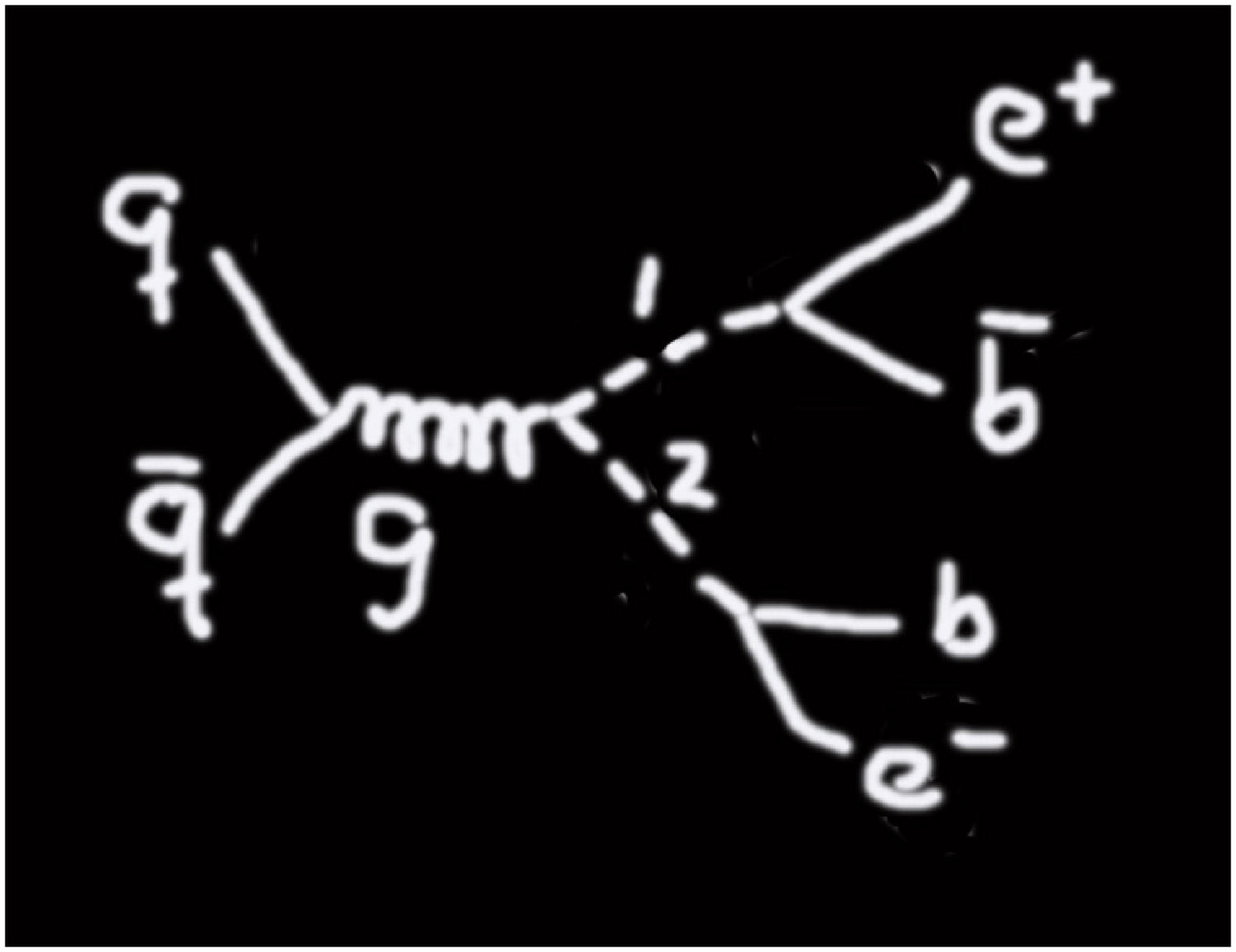} \\
\end{tabular}
\caption{Chalkboard drawing of the ingoing and outgoing legs of the Feynman diagram responsible for producing an observed signal in the final state $e^+e^-b\bar{b}$ at the Tevatron (left), and of a Feynman diagram possibly responsible for producing this signal (right).\label{fig:eebbChalkboard}}
\end{figure}

 In contemporary high energy physics experiments, it is not uncommon to observe discrepancies between data and Standard Model predictions.  Most of these discrepancies have been explained away over time. To convincingly demonstrate that an observed effect is evidence of physics beyond the Standard Model, it is necessary to prove it is (1) not a likely statistical fluctuation, (2) not introduced by an imperfect understanding of the experimental apparatus, (3) not due to an inadequacy of the implementation of the Standard Model prediction, and (4) interpretable in terms of a sensible underlying theory.  Those who object to (4) as being necessary fail to appreciate that most hypothesis development in science occurs before, rather than after, publication.  This last criterion is essential, and will likely point the way to other discrepancies that must exist if the interpretation is correct.

In the search for new electroweak-scale physics at the frontier energy colliders, a model-independent search strategy
(\Vista~\cite{ACAT2003:Knuteson:2004nj,Moriond2005:Knuteson:2005ev}
or \Sleuth~\cite{ACAT2003:Knuteson:2004nj,Moriond2005:Knuteson:2005ev,
SleuthPRL:Abbott:2001ke,SleuthPRD1:Abbott:2000fb,SleuthPRD2:Abbott:2000gx})
rigorously addresses whether a statistical fluctuation explains the
 observation. Rejecting the hypothesis that the observed effect arises
from
 a feature of the detector or an inadequacy of the detector simulation
is best
 handled by requiring consistency among all collected data; this is the
purpose of \Vista.
 Our ability to calculate QCD at hadron colliders has improved
dramatically over the past decade, with much recent progress in describing multi-jet
final states.  Using these tools and demanding consistency among
many different observables addresses the third
criterion.
 Addressing the fourth requires a practical method for
systematically generating new hypotheses to yield sensible interpretations of discrepancies. 
%

Event generators containing implementations of physics beyond the Standard Model are able to calculate model predictions within particular scenarios.  Interpreting a specific discrepancy requires working in the inverse direction, from observed phenomenon to the underlying model.  The typical top-down approach of scanning model parameter spaces to find regions compatible with discrepancies is computationally intractable for parameter spaces with dimensionality larger than about five.  
We are aware of no satisfactory systematic prescription for interpreting possible discrepancies observed at the Tevatron or Large Hadron Collider in terms of the new underlying physics. This Letter introduces \Bard, a bottom-up algorithm whose function is to weave a story to explain observation.

 Working in an effective field theoretic framework, we write ${\cal
L}_{\cal H} = {\cal L}_{\text{SM}} + {\cal L}_{\text{new}}$, where
${\cal H}$ denotes a new hypothesis, the sum of Standard Model
Lagrangian terms ${\cal L}_{\text{SM}}$ and new terms ${\cal
L}_{\text{new}}$ entailing additional Feynman diagrams. Our goal is to
determine what new term(s) ${\cal L}_{\text{new}}$ best describe a
particular observed discrepancy in the data.
The ability to generate new predictions automatically is
facilitated by progress in the calculation of the Standard Model: 
\MadEvent~\cite{MadEvent:Maltoni:2002qb} and other tools are able to provide the 
Standard Model prediction exactly at tree level for arbitrary final 
states of low multiplicity, and other efforts are pushing systematic 
calculations to one loop.

The result of \Vista\ or \Sleuth\ is a discrepancy observed in a
particular final state, perhaps on the tail of the distribution of the total
summed scalar transverse momentum in the event, as pictured in cartoon
form in Fig.~\ref{fig:eebbSleuthCartoon}. In determining the Feynman
diagram(s) potentially responsible for producing the observed effect,
the nature of the incoming particles determines the incoming legs in the
graphs of interest, and the particular final state in which the
discrepancy is observed determines the outgoing legs. This is shown as a
chalkboard drawing in Fig.~\ref{fig:eebbChalkboard}(a). The game is to
provide the middle part of the graph, such as shown in
Fig.\ref{fig:eebbChalkboard}(b).

\Bard\ begins by exhaustively listing reasonable possibilities, involving all operators with mass dimension four or less, and introducing generic new particles of spin 0, 1/2, or 1; having electric charge in multiples of 1/3; and existing as singlets, triplets, or octets under SU(3)$_\text{color}$.

\Bard\ uses \MadGraph~\cite{MadGraph:Stelzer:1994ta} to systematically generate all diagrams entailed by these new terms, an example of which is shown in Fig.~\ref{fig:eebbChalkboard}(b).  No attention is paid at this stage to whether the particles and interactions introduced fit naturally into a fashionable theoretical framework.  The resulting diagrams are partitioned into stories, collections of diagrams in which the existence of any single diagram in the story implies the existence of the others.  Depending on the final state, \Bard\ will generate between a few and a few thousand stories as potential explanations for the observed discrepancy.

 Each story introduces several new parameters. These parameters are the
masses and widths of the introduced particles, and the couplings at each
vertex. This parameter space is sufficiently small that it can be
scanned, provided a fast yet sensitive analysis algorithm exists to test
each of these stories as an explanation for the observed effect.
\Quaero~\cite{QuaeroPRL:Abazov:2001ny,chep2003Quaero:Knuteson:2003dn}
was designed for this purpose.

\Bard\ passes the new Lagrangian terms ${\cal L}_{\text{new}}$ to \Quaero, which has been prepared with the interesting subset of the data highlighted by \Vista\ or \Sleuth.  \Quaero\ uses \MadEvent\ to integrate the squared amplitude over the available phase space and to generate representative events, and uses \Pythia~\cite{Pythia:Sjostrand:2000wi} for the showering and fragmentation of these events.  \TurboSim\ is used as a fast replacement for the experiment's full detector simulation.  \Quaero\ performs the analysis, numerically integrating over systematic errors, returning as output $\log_{10}{\cal L}$, where ${\cal L} = p({\cal D}|{\cal H})/p({\cal D}|\text{SM})$ is a likelihood ratio, representing the probability of observing the data ${\cal D}$ assuming the hypothesis ${\cal H}$ divided by the probability of observing the data ${\cal D}$ assuming the Standard Model alone.  The region in the parameter space of the story that maximizes $\log_{10}{\cal L}$ is determined, providing also an error estimate on the parameter values.  Repeating this process in parallel for each story enables an ordering of the stories according to decreasing goodness of fit to the data.

 The testing discussed so far occurs only on that subset of data in
which the discrepancy is observed. Once the list of stories has been
ordered, those at the top of the list can be tested further. In the
example provided in Fig.~\ref{fig:eebbChalkboard}, a story involving a
$Z$ boson as an intermediate state decaying to $e^+e^-$ must produce
effects also in $\mu^+\mu^-b\bar{b}$ and $\tau^+\tau^-b\bar{b}$.  A story involving the pair production of charge $4/3$ leptoquarks coupling the first lepton generation with the third quark generation might (by crossing) have other observable consequences at LEP or HERA, depending on the leptoquark mass.
The broader consequences of the most compelling stories can then be worked
out against all frontier energy collider data using \Quaero.

 Simplifications to the procedure described above decrease the
computational cost of the algorithm. Vectors and scalars enter in
similar ways into the stories considered; either spin 0 or spin 1
particles can be discarded. Electric and color charge and fermion number
conservation may be assumed at each vertex. Vertices with four external
legs can be ignored. When generating the list of diagrams, it is
convenient to exclude those diagrams containing propagators that are not
new particles, the top quark, or a gauge boson, on the grounds that
a diagram involving a light internal propagator would likely first appear
as a discrepancy in another final state through the subdiagram obtained by cutting through the light internal propagator. The widths of the particles can
be taken to be small compared to experimental resolution. Since the
couplings of diagrams in each story enter only as the square of their
product, the parameters associated with each story are one mass for each
new particle added, and one overall coupling; this parameter space is
most efficiently explored by scanning in the subparameter space of
masses, and for each choice of particle masses exploiting the known
shape of $\log_{10}{\cal L}$ as a function of the overall coupling to
find the maximum. Final states with missing energy require a loop over
neutrinos and heavy new particles lacking strong and electromagnetic
interactions.  Interference between Standard Model and new diagrams can be ignored.  Stories involving only one new particle may first be considered, and stories involving two or three new particles considered secondarily.  Assumptions such as these explicitly limit the story space in the interest of speed.


Starting bottom-up from a specific observed discrepancy, \Bard\ is able to perform a more targeted search than those who scan model parameter spaces.  \Bard\ will allow an experiment to publish an observed discrepancy together with an extensive list of possible interpretations, with this list ordered according to how well each story fits the data, and with best fit parameter values for each story.  Multiple discrepancies are naturally handled sequentially by \Bard.  A systematic approach will likely be required in sorting out scenarios involving a complex spectrum of new resonances, such as supersymmetry, with \Bard\ regularly suggesting possible explanations of the data that might otherwise be overlooked for years.  As an unanticipated advantage, \Bard\ is also able to determine whether an observed discrepancy has any possible underlying interpretation at all, and assists in understanding which of our assumptions must be violated for an underlying interpretation to exist.

The new theory ${\cal L}_{\cal H}$ is at this point the Standard Model Lagrangian ${\cal L}_{\text{SM}}$ patched with additional terms ${\cal L}_{\text{new}}$ to explain particular effects.  There will likely be no practical possibility to divine a deeper structure until several such additional terms have been added to explain several discrepancies.  Once several such new terms have been added, deriving the deeper structure is largely a matter of identifying similar terms in ${\cal L}_{\cal H}$, and writing the Lagrangian more compactly.  If the $W$ and $Z$ bosons, the top quark, and the Higgs boson were not already known, one could imagine deducing the Standard Model from LEP, Tevatron, and future LHC data in this manner.

 We expect the systematic, bottom-up approach encapsulated in the \Bard\ algorithm and described in this Letter to be useful for interpreting impending discoveries at the Tevatron and Large Hadron Collider.  In the problem domain of interpreting new electroweak scale physics from the current generation of frontier energy colliders, the details of the algorithm are sufficiently worked out to be reasonably confident of its success.  More generally, the spirit of automatic model construction described here has application to other interpretations of data that take the form of an effective Lagrangian.  In these problem domains the details of a workable algorithm may or may not turn out to be as trivial as we have found them to be at the electroweak scale.  More generally still, the systematization of model construction may eventually play a useful role in other subfields of science.

\acknowledgments

Tim Stelzer (UIUC) provided \MadGraph\ and \MadEvent, two crucial ingredients in the approach advocated in this Letter.  Conversations with Michael Niczyporuk (MIT) led to the congealing of the ideas described here.  Khaldoun Makhoul and Georgios Choudalakis (MIT) assisted in \Bard's implementation.  Financial support for this effort comes in part from a Department of Defense Graduate Science and Engineering Fellowship at the University of California at Berkeley; NSF International Research Fellowship INT-0107322 at CERN; a Fermi/McCormick Fellowship at the University of Chicago; and DoE grant DE-FC02-94ER40818 at the Laboratory for Nuclear Science at MIT.

\bibliography{prl}

\begin{thebibliography}{10}
\expandafter\ifx\csname natexlab\endcsname\relax\def\natexlab#1{#1}\fi
\expandafter\ifx\csname bibnamefont\endcsname\relax
  \def\bibnamefont#1{#1}\fi
\expandafter\ifx\csname bibfnamefont\endcsname\relax
  \def\bibfnamefont#1{#1}\fi
\expandafter\ifx\csname citenamefont\endcsname\relax
  \def\citenamefont#1{#1}\fi
\expandafter\ifx\csname url\endcsname\relax
  \def\url#1{\texttt{#1}}\fi
\expandafter\ifx\csname urlprefix\endcsname\relax\def\urlprefix{URL }\fi
\providecommand{\bibinfo}[2]{#2}
\providecommand{\eprint}[2][]{\url{#2}}

\bibitem[{\citenamefont{Knuteson}(2004)}]{ACAT2003:Knuteson:2004nj}
\bibinfo{author}{\bibfnamefont{B.}~\bibnamefont{Knuteson}},
  \bibinfo{journal}{Nucl. Instrum. Meth.} \textbf{\bibinfo{volume}{A534}},
  \bibinfo{pages}{7} (\bibinfo{year}{2004}), \eprint{hep-ex/0402029}.

\bibitem[{\citenamefont{Knuteson}(2005)}]{Moriond2005:Knuteson:2005ev}
\bibinfo{author}{\bibfnamefont{B.}~\bibnamefont{Knuteson}}
  (\bibinfo{year}{2005}), \eprint{hep-ex/0504041}.

\bibitem[{\citenamefont{Abbott
  et~al.}(2001{\natexlab{a}})}]{SleuthPRL:Abbott:2001ke}
\bibinfo{author}{\bibfnamefont{B.}~\bibnamefont{Abbott}} \bibnamefont{et~al.}
  (\bibinfo{collaboration}{D0}), \bibinfo{journal}{Phys. Rev. Lett.}
  \textbf{\bibinfo{volume}{86}}, \bibinfo{pages}{3712}
  (\bibinfo{year}{2001}{\natexlab{a}}), \eprint{hep-ex/0011071}.

\bibitem[{\citenamefont{Abbott et~al.}(2000)}]{SleuthPRD1:Abbott:2000fb}
\bibinfo{author}{\bibfnamefont{B.}~\bibnamefont{Abbott}} \bibnamefont{et~al.}
  (\bibinfo{collaboration}{D0}), \bibinfo{journal}{Phys. Rev.}
  \textbf{\bibinfo{volume}{D62}}, \bibinfo{pages}{092004}
  (\bibinfo{year}{2000}), \eprint{hep-ex/0006011}.

\bibitem[{\citenamefont{Abbott
  et~al.}(2001{\natexlab{b}})}]{SleuthPRD2:Abbott:2000gx}
\bibinfo{author}{\bibfnamefont{B.}~\bibnamefont{Abbott}} \bibnamefont{et~al.}
  (\bibinfo{collaboration}{D0}), \bibinfo{journal}{Phys. Rev.}
  \textbf{\bibinfo{volume}{D64}}, \bibinfo{pages}{012004}
  (\bibinfo{year}{2001}{\natexlab{b}}), \eprint{hep-ex/0011067}.

\bibitem[{\citenamefont{Maltoni and Stelzer}(2003)}]{MadEvent:Maltoni:2002qb}
\bibinfo{author}{\bibfnamefont{F.}~\bibnamefont{Maltoni}} \bibnamefont{and}
  \bibinfo{author}{\bibfnamefont{T.}~\bibnamefont{Stelzer}},
  \bibinfo{journal}{JHEP} \textbf{\bibinfo{volume}{02}}, \bibinfo{pages}{027}
  (\bibinfo{year}{2003}), \eprint{hep-ph/0208156}.

\bibitem[{\citenamefont{Stelzer and Long}(1994)}]{MadGraph:Stelzer:1994ta}
\bibinfo{author}{\bibfnamefont{T.}~\bibnamefont{Stelzer}} \bibnamefont{and}
  \bibinfo{author}{\bibfnamefont{W.~F.} \bibnamefont{Long}},
  \bibinfo{journal}{Comput. Phys. Commun.} \textbf{\bibinfo{volume}{81}},
  \bibinfo{pages}{357} (\bibinfo{year}{1994}), \eprint{hep-ph/9401258}.

\bibitem[{\citenamefont{Abazov et~al.}(2001)}]{QuaeroPRL:Abazov:2001ny}
\bibinfo{author}{\bibfnamefont{V.~M.} \bibnamefont{Abazov}}
  \bibnamefont{et~al.} (\bibinfo{collaboration}{D0}), \bibinfo{journal}{Phys.
  Rev. Lett.} \textbf{\bibinfo{volume}{87}}, \bibinfo{pages}{231801}
  (\bibinfo{year}{2001}), \eprint{hep-ex/0106039}.

\bibitem[{\citenamefont{Knuteson}(2003)}]{chep2003Quaero:Knuteson:2003dn}
\bibinfo{author}{\bibfnamefont{B.}~\bibnamefont{Knuteson}},
  \bibinfo{journal}{ECONF} \textbf{\bibinfo{volume}{C0303241}},
  \bibinfo{pages}{TULT001} (\bibinfo{year}{2003}), \eprint{hep-ex/0305065}.

\bibitem[{\citenamefont{Sjostrand et~al.}(2001)}]{Pythia:Sjostrand:2000wi}
\bibinfo{author}{\bibfnamefont{T.}~\bibnamefont{Sjostrand}}
  \bibnamefont{et~al.}, \bibinfo{journal}{Comput. Phys. Commun.}
  \textbf{\bibinfo{volume}{135}}, \bibinfo{pages}{238} (\bibinfo{year}{2001}),
  \eprint{hep-ph/0010017}.

\end{thebibliography}

\end{document}